\documentclass[12pt]{iopart}
\usepackage{amsmath}
\usepackage{hyperref}
\usepackage{braket}
\usepackage{tikz}
\usepackage{cleveref}
\usepackage{comment}
\usepackage{tikz}
\usepackage{cite}
\newcommand{\Plus}{\mathord{\begin{tikzpicture}[baseline=0ex, line width=1, scale=0.13]
\draw (1,0) -- (1,2);
\draw (0,1) -- (2,1);
\end{tikzpicture}}}

\definecolor{byzantine}{rgb}{0.74, 0.2, 0.64}

\begin{document}

\title{Purifying quantum-dot light in a coherent frequency interface}

\author{F Chiriano, C L Morrison, J Ho, T Jaeken and A Fedrizzi}
\address{Institute of Photonics and Quantum Sciences, School of Engineering and Physical Sciences, Heriot-Watt University, Edinburgh EH14 4AS, United Kingdom}
\ead{fc2005@hw.ac.uk}
\vspace{10pt}
\begin{indented}
\item[]April 2024 
\end{indented}

\begin{abstract}
Quantum networks typically operate in the telecom wavelengths to take advantage of low-loss transmission in optical fibres. However, bright quantum dots (QDs) emitting highly indistinguishable quantum states of light, such as InGaAs QDs, often emit photons in the near infrared thus necessitating frequency conversion (FC) to the telecom band. Furthermore, the signal quality of quantum emissions is crucial for the effective performance of these networks. In this work we report a method for simultaneously implementing spectral purification and frequency shifting of single photons from QD sources to the C-band in a periodically poled Lithium Niobate waveguide. We consider difference frequency generation in the counter-propagating configuration to implement FC with the output emission bandwidth in units of GHz. Our approach establishes a clear path to integrating high-performance single-emitter sources in a hybrid quantum network.
\end{abstract}

\section{Introduction}
Quantum light sources power a range of applications in quantum technology, in particular quantum networks.
Single-photon sources suitable for networks have to deliver high generation rates (brightness), with a high probability of an emission per clock cycle.
They have to produce photons with high photon-number and spectral purity for applications based on multi-photon interference.
They have to be compatible with the telecom transmission window at 1550~nm and finally be reasonably narrow-band to reduce dispersion and for addressing receivers such as quantum memories efficiently~\cite{Meyer_2020}. 

Heralded photon pair sources based on nonlinear processes such as parametric down-conversion~\cite{Pickston_2021} or four-wave mixing~\cite{Spring2017} are attractive options as they offer high generation rates, near-unity spectral purity and operate at room temperature.
However, they tend to be spectrally broad (100s of GHz) and photon-number purity decreases with increasing brightness~\cite{Migdall_2002,Pittman2002}, typically requiring multiplexing~\cite{Christ_2012,Brida_2012}, e.g., temporally~\cite{Broome_2011,xiong2016},  spatially~\cite{ma2011}, or both~\cite{Mendoza2016}, to achieve high photon-number purity.
Conversely, single quantum emitters such as quantum dots (QDs) are spectrally narrow-band and they can achieve high photon-number purity due to not relying on probabilistic photon creation.
They do however lag behind in spectral purity---photons from dissimilar dots have little spectral overlap without active tuning, and even photons emitted from the same dot tend to be distinguishable due to environmental factors, including phonon interactions, charge noise and blinking~\cite{yu2023naturenano}.
This has led to resource-demanding techniques for purifying QD photons~\cite{gao2019physrevlett,faurby2024}.
Furthermore, despite rapid advances with dots emitting in the telecom C-band~\cite{Muller_2018,Nawrath_2019,dusanowski2022,nawrath2023,Phillips_2024,holewa2024high,Thomas2024sciadv}, the quantum dots with the best metrics remain those based on InAs/GaAs, which tend to emit in the region of 920--980~nm~\cite{Tomm_2021,Cao_2024}.

\begin{figure}[tb]
\centering
\includegraphics[width = \columnwidth]{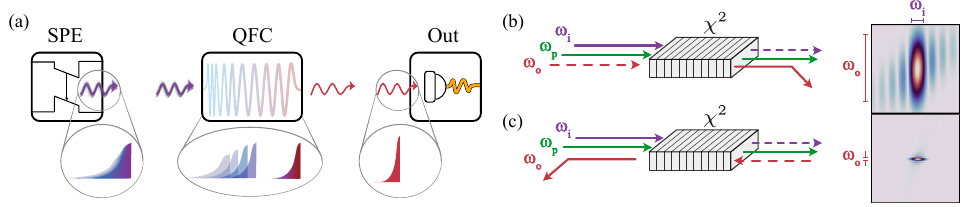}
\caption{Outline of the scheme.
(a) Light from single photon emitter (SPE) undergoes quantum FC into the telecom band. We tailor the FC to avoid spectral broadening of the output spectrum and purification of the output state exploiting three-wave mixing processes, i.e., materials with $\chi^{(2)}$ non-linearity. (b) DFG process in co-propagation configuration, with characteristic transfer function given by joint spectral function of the input ($\omega_i$) and output ($\omega_o$) fields. (c) In the reverse-wave DFG process, the output field propagates in the opposite direction to both the input and pump fields ($\omega_p$), leading to a narrow output spectral function.}
\label{Fig: conceptual_figure}
\end{figure}

Here we propose a coherent quantum interface which spectrally purifies QD photons while simultaneously converting them from the short-infrared regime to the telecom C-band via difference-frequency generation (DFG), see \cref{Fig: conceptual_figure}(a). Quantum frequency conversion has successfully been demonstrated to convert the best performing QDs from $\sim$900~nm to telecom wavelengths $\sim$1550~nm with near unity internal efficiencies~\cite{pelc2012,Kambs_16,Weber_2019,morrison2021apl,da2022,morrison2023single}. 
In our scheme, we use the quantum pulse gate (QPG) toolbox to selectively convert the fundamental spectral mode of a photon while suppressing higher-order modes~\cite{Eckstein2011opticsexpress,Brecht_2011} of a spectrally mixed input photon.
Our approach is inspired by a similar scheme where this is achieved in a quantum buffer~\cite{gao2019physrevlett}.
In DFG, this is possible in the so-called asymmetric group velocity matching (aGVM) regime, where the pump laser group velocity is matched to either the input or the output photon, for a suitable set of wavelengths~\cite{URen2005LasPhys}.
We will first outline the theory of spectral purification in DFG and then consider a practical example for a very common scenario ~\cite{Chatterjee2021naturrevphys,yu2023naturenano} in which aGVM already holds through serendipity: the  conversion of a 942~nm photon from an InGaAs QD to 1550~nm using DFG driven by a 2.4~$\mu$m seed laser in a PPLN waveguide.
The mode selectivity through aGVM frequency conversion is well established in the forward-pass configuration~\cite{Serino2023PRX}, where all three fields co-propagate, but it typically results in broadened spetcral outputs.
Thus, we phasematch our waveguide for the counter-propagating photon configuration ensuring the daughter photon to have significantly narrower spectral width of just a few GHz.
We simulate purification by artificially adding time and frequency noise to the QD photon~\cite{gao2019physrevlett} and show that we can achieve single-mode frequency conversion with near-unity conversion efficiency while effectively eliminating mixture.
More importantly, we show that our technique not only outperforms passive spectral filters in removing frequency jitter but is also uniquely capable of eliminating time jitter, a feat that passive filters cannot achieve~\cite{Raymer_2020}.

\section{Method}

In DFG, an input field at frequency $\omega_i$ converts to a lower-energy output field at frequency $\omega_o$ assisted by a strong pump field at frequency $\omega_p = \omega_i - \omega_o$.
In a first-order approximation, the DFG output state is expressed as~\cite{Christ2013iop}
\begin{equation}
\ket{\psi} = \iint \text{d}\omega_o\text{d}\omega_i F\left(\omega_i,\omega_o\right) g\left(\omega_i\right)\hat{a}^{\dagger}\left(\omega_o\right)\ket{0},
\label{eq:dfg_outputState1}
\end{equation}
where $g\left(\omega_i\right)$ is the spectral envelope of the input field, which in our case is a QD photon.
A key role in describing the spectral structure of the state is played by the joint spectral distribution (JSD) $F\left(\omega_i,\omega_o\right)$~\cite{Brecht_2011}.
This function acts as the energy-time mapping equivalent of the joint spectral amplitude (JSA)~\cite{Graffitti_2018} in PDC processes.
Unlike the JSA though,
the JSD describes the transformation of input state modes into the output state, effectively serving as a transfer function for frequency conversion processes~\cite{Ansari_2018}. 
Like in PDC, the JSD is defined as the product of the pump envelope function (PEF) treated as a classical field $\alpha(\omega_i,\omega_o)$, and the phase matching function (PMF) $\phi(\omega_i,\omega_o)$:
\begin{equation}
    F(\omega_i,\omega_o) = \frac{1}{\mathcal{N}}\alpha(\omega_i,\omega_o) \phi(\omega_i,\omega_o),
\label{eq:JSA}
\end{equation}
with the normalization constant  $\mathcal{N}$~\cite{Brecht_2011}.
The PEF represents the spectral properties of the pump and governs the energy conservation during conversion. For simplicity, we assume $\alpha(\omega_i,\omega_o)$ to be Gaussian.
The PMF $\phi(\omega_i,\omega_o)$ defines the region in which momentum is conserved during the process and depends on the dispersive properties of the non-linear material.

Here we consider a periodically poled waveguide, which simplifies the PMF to~\cite{Christ2013iop}:
\begin{equation}
    \phi(\omega_i,\omega_o) = \text{sinc}\left(\frac{\Delta k \cdot L}{2}\right).
\label{eq:pmf}
\end{equation}
In the standard DFG configuration, the three waves co-propagate during the process.
Spectral purification is achievable in this regime~\cite{Brecht_2011,Serino2023PRX}, however, it comes at the price of drastically broadening the typically very narrow spectrum of QD photons, see~\cref{Fig: conceptual_figure}(b).
We will therefore model our purification DFG interface for the counter-propagating photon configuration, where the output photon propagates in the opposite direction to the pump and the input photon. The phase mismatch term in this scenario is \(\Delta k = k_i - k_p \Plus k_o = \frac{2m\pi}{\Lambda}\).
This arrangement significantly narrows the acceptance bandwidth of the JSD, which is illustrated in~\cref{Fig: conceptual_figure}(c). 
The pump and output signal in this scenario have virtually identical group velocities, which results in a horizontally aligned, separable JSD. 
This ensures that the input and output are effectively uncorrelated, which allows the generation of a perfectly pure output photon in the conversion process~\cite{Ansari_2018}.

To explain the concept of our purification scheme, we use the Schmidt decomposition~\cite{Ansari_2018} to express the JSD with two sets of complete orthogonal bases, $g^{(i)}$ and $h^{(o)}$:
\begin{equation}
    F(\omega_i,\omega_o)= \sum_{j=1}^{\infty} \sqrt{d_j} g^{(i)}_j(\omega_i) h^{(o)}_j(\omega_o).
    \label{eq:JSA_shmidt_deco}
\end{equation}
Each mode $j$ corresponds to a positive and real Schmidt coefficient $\sqrt{d_j}$, which satisfy \emph{completeness} $\sum_{j=0}^{\infty} d_j = 1$. We call the inverse of the summation of all the coefficients $d_j$ Schmidt number of the JSD, $K_{\text{JSD}}=1/\sum_{j=0}^{\infty} d^2_j$.
Although \cref{eq:JSA_shmidt_deco} appears to be in the same form for both FC and PDC, there is a crucial distinction between the interpretation of the two.
Whereas the PDC decomposition reveals the longitudinal modes of a quantum state, for FC it defines mapping operators from the input to the output state~\cite{raymer2010optcomm,brecht2015physrevx,Ansari_2018}. 
More specifically, one can define a transmission operator and a reflection operator to link the FC process to a two-color beam-splitter operation~\cite{raymer2010optcomm}.
In this analogy, each input mode $g_j^{(i)}$ is either reflected into the output mode $h_j^{(o)}$ or transmitted as $g_j^{(o)}$, determining whether the input state in mode $j$ undergoes frequency conversion or not.
In this picture, the Schmidt number of the JSD indicates the number of modes allowed in the conversion.

The conversion occurs with efficiency
\begin{equation}
    \eta_j = \sin^2{\theta_j},
\label{eta}
\end{equation}
where $\theta_j=\sqrt{d_j} \theta$ is the coupling constant associated to mode $j$.
The constant $\theta$ can be interpreted as the extinction ratio of the two-color beam-splitter and it is a function of both the properties of the interacting fields and the dispersive medium.
Its full expression is derived in~\cite{Brecht_2011} by integrating the FC Hamiltonian over time and space and reads as follows,
\begin{equation}
    \theta = \frac{2d_{\text{eff}}\pi^2L\mathcal{N}}{c}\sqrt{\frac{2\omega_{\text{i}}\omega_{\text{o}}}{c\epsilon_0 n_{\text{p}} n_{\text{i}} n_{\text{o}} |\int d\omega_{\text{p}} \alpha(\omega_{\text{p}})|^2}} \sqrt{\frac{P_{\text{p}}}{\mathcal{A}_{\text{eff}}}}.
\label{eq:theta}
\end{equation}
The coupling constant $\theta$ includes several key parameters that contribute to the conversion efficiency: 
the length of the medium, $L$, which also affects the spectral width, as we show in the next section; the effective non-linearity $d_{\text{eff}}$ of the material;
the electric field energy density $|\int d\omega_{\text{p}} \alpha(\omega_{\text{p}})|^2$; the cross section of the interaction, $\mathcal{A_{\text{eff}}}$; and the material's refractive indices at the wavelengths of the interacting fields—$n_{\text{p}}$, $n_{\text{i}}$, and $n_{\text{o}}$. 

To purify an input photon with spectro-temporal noise, we need to optimize the conversion process for a single spectral mode and simultaneously suppress remaining modes via fine-tuning of the parameters in \cref{eq:theta}.
Specifically, we determine when the coupling constant $\theta$ approximates a single-mode converter in the JSD.
Since the number of modes converted by this process is effectively given by its Schmidt number $K_{\text{JSD}}$, 
in the ideal case $K_{\text{JSD}} = 1$ then the JSD becomes fully separable which ensures a spectrally pure output state.
Remarkably, this process can achieve unit conversion efficiency of the first Schmidt mode, i.e. the zeroth mode, because the JSD selectively, and completely, converts photons in this mode.
The JSD therefore acts in every respect as a spectral filter in amplitude, in contrast with the intensity filters which are commonly used after FC to discard photons outside a certain frequency range, which will also remove some photons in the desired zeroth mode.

\begin{figure}
    \includegraphics[width=\linewidth]{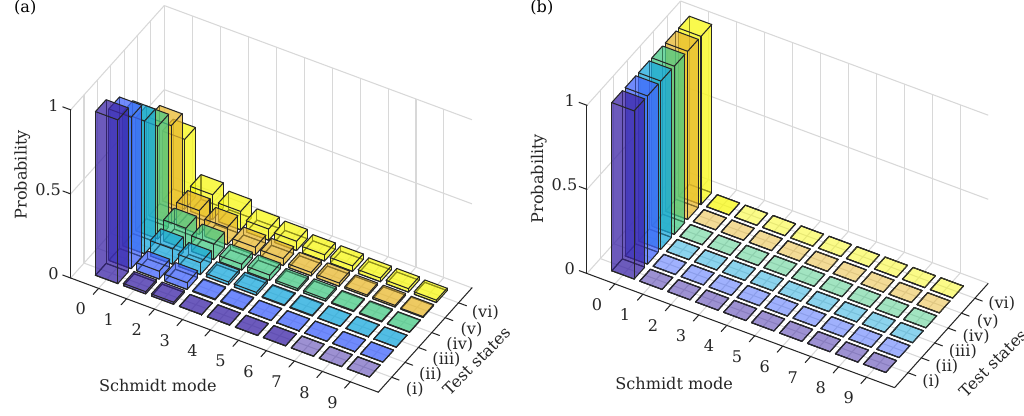}
    \caption{Schmidt decomposition of (a) initial and (b) final quantum signals for a set of test states (i)-(vi), which have decreasing initial state purity from added frequency-jitter noise.
    In each case, the frequency conversion process is optimized by setting the power to achieve unit conversion efficiency of the zeroth mode, while higher-order modes are suppressed.}
    \label{Fig: initial_vs_finalPurityBars}
\end{figure}

Conceptually, our spectral purification process is shown in \cref{Fig: initial_vs_finalPurityBars}.
Here we simulate a set of initial states (i)-(vi), with decreasing purity by adding frequency noise on the QD single photon as in~\cite{gao2019physrevlett}.
We write the QD photon in the density matrix formalism $\rho_{\text{i}}$ and we define $p(t_0,\omega_0)$ as the probability distribution describing any spectro-temporal broadening~\cite{gao2019physrevlett},
\begin{equation}
    \rho = \int \text{d}t_0 \text{d}\omega_0 p(t_0,\omega_0) \rho_{\text{i}}(t_0,\omega_0).
\label{eq:output_density_matrix}
\end{equation}
In our model, the distribution $p(t_0,\omega_0)$ is Gaussian in both time and frequency domain.
Injecting noise in the state leads to a gradual increase in the contributions of higher order modes, making the it more multi-modal, as in ~\cref{Fig: initial_vs_finalPurityBars}(a).
We then apply our technique to each state, optimising the conversion efficiency for the zeroth mode.
After the conversion, we renormalize the output state such that its trace over the Schmidt modes is equal to 1.
As a result, the energy-time structure of the output state becomes completely single mode and pure, as depicted in~\cref{Fig: initial_vs_finalPurityBars}(b).

In the following section we will apply this frequency-conversion purification scheme to the specific case of a 942~nm QD photon converted to 1550~nm in a periodically-poled lithium niobate (PPLN) waveguide phase-matched for counter-propagating input and output fields.

\section{Results}
The parameters available to optimise purification while also maintaining a high system frequency conversion efficiency for a given set of input and output wavelengths are the waveguide length, the pump pulse duration, and the pump peak power. 
We begin by examining the impact of the waveguide length $L$ on the zero-mode conversion efficiency, normalised at its maximum, relative to peak power requirements, which decreases with increased $L$. 
Next, we identify the optimal pump pulse duration that minimizes $K_{\text{JSD}}$ for a given $L$.
We will then examine the achievable purity and normalized conversion efficiency of the emitted photon, and finally highlight the advantages of operating in the counter-propagating regime to avoid the spectral broadening seen in co-propagating setups.

Inspired by a series of experiments in this regime~\cite{morrison2021apl,morrison2023single}, 
we will study a type-0 collinear DFG process, driven by a 2.4~$\mu\text{m}$ pulsed laser with a Gaussian spectral envelope. 
This setup facilitates the conversion of a single photon emitted by an InGaAs QD at 942~nm into a 1550~nm signal within a PPLN ridge waveguide. 
The waveguide transverse dimensions are $13\mu\text{m} \times 13.4\mu\text{m}$, which is wide enough, compared to the three interacting fields, to neglect waveguide dispersion and use the bulk crystal approximation.
As previously mentioned, the PPLN crystal we model is phase matched in the counter-propagating regime, which for these wavelengths requires a zero-order poling period of $\sim 360$~nm.

\begin{figure}[!h]
    \includegraphics[width=\linewidth]{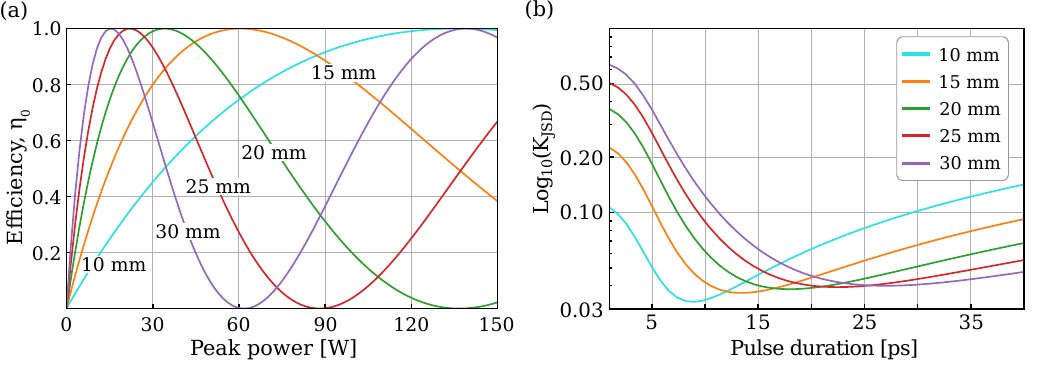}
    \caption{Visualization of key experimental parameters affecting conversion efficiency and photon purity.
    (a) The conversion efficiency of the zeroth mode normalised at its maximum, $\eta_{0}$, versus peak power of the pump laser for different waveguide lengths.
    (b) The Schmidt number versus pulse duration for different waveguide lengths.
    }
    \label{Fig: crystal_length_pulse_duration}
\end{figure}

In \cref{Fig: crystal_length_pulse_duration}(a) we plot the zeroth mode conversion efficiency normalised at is maximum, $\eta_0$, as a function of peak power for several different waveguide lengths, $L$, using \cref{eq:theta}.
As an example, for the $15$~mm waveguide we find that peak power required for unit conversion efficiency of the zeroth mode is $\sim$60~W.
All else being equal, as the waveguide length increases the peak power required to achieve unit conversion efficiency of the zeroth mode reduces.
As such, using longer waveguides may be favourable in setups with limited pump laser power overhead however, as we show next, longer waveguides reduce the attainable spectral purities.

In \cref{Fig: crystal_length_pulse_duration}(b) we plot the Schmidt number $K_{\text{JSD}}$ as a function of pulse duration for different waveguide lengths, $L$.
The pulse duration is defined for a transform-limited pump laser, which we convert to the spectral bandwidth of a Gaussian PEF, $\alpha(\omega_i, \omega_o)$.
To arrive at the Schmidt number, we use \cref{eq:JSA} satisfying the phase mismatch $\Delta k$ for the counter-propagating regime, then obtain the Schmidt decomposition in the form of \cref{eq:JSA_shmidt_deco}.
Ideally, $K_{\text{JSD}}$ should be as close to $1$ as possible for conversion of a single spectral mode, which is required for purifying the spectral modes of the output photon.
We observe that longer waveguide lengths, $L$, require longer pump pulse duration to reach the minimum Schmidt number values, this in turn requires higher average optical power of the pump laser.
For the 15~mm waveguide we selected earlier, the optimal pulse duration is $\sim$13~ps, resulting in $K_{\text{JSD}}=1.037$.
From an operational viewpoint, a pulsed $2.4~\mu$m laser with a pulse duration of 13~ps, repetition rate of 80~MHz, and a peak power of $\sim$60~W corresponds to an average optical power output of $\sim$62~mW which is achievable with commercial optical parametric oscillators.
Lastly, we observe that for longer waveguide lengths, $L$, the minimum optimizable value of the Schmidt number increases, which translates to a lower purity.

\begin{figure}[!h]
    \centering
\includegraphics[width=\linewidth]{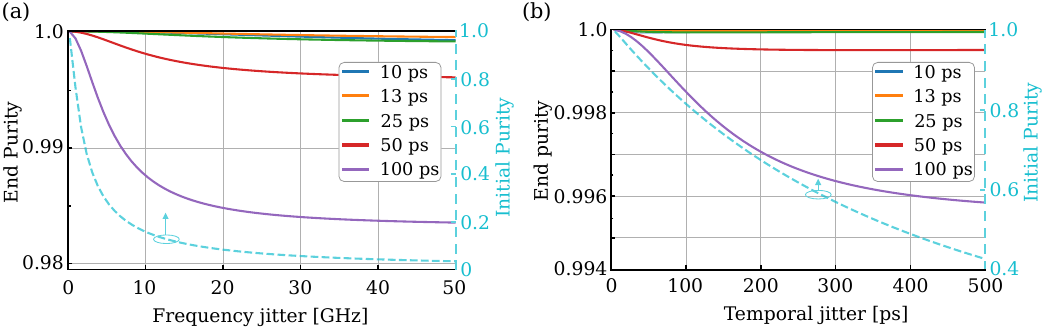}
    \caption{(a) Solid lines are the achievable final purity as a function of injected frequency noise on the QD single photon, for different pump pulse duration. Dashed line is the initial purity owing to the noise, axis on the right. (b) A similar analysis is performed using temporal jitter as the noise.}
    \label{fig:purity_with_freq_time_noise}
\end{figure}

In \cref{fig:purity_with_freq_time_noise} we evaluate the achievable performance of our scheme in the presence of frequency and temporal noise applied to the QD photon, as in~\cref{eq:output_density_matrix}.
Here, we simulated a range of noisy states ranging from zero noise up to 50~GHz of frequency jitter, in \cref{fig:purity_with_freq_time_noise}(a), and up to 500~ps for temporal noise, in \cref{fig:purity_with_freq_time_noise}(b).
For each noise value, we calculate the corresponding initial purity of the state.
We then apply our scheme for a range of different pump durations and calculated the output mode purity, keeping the waveguide length $L=15$ mm.
We observe that our technique is able to purify the inputs even if the pump duration is not optimal. 
As expected, the optimal trend corresponds to the optimized pump duration for the chosen waveguide length, i.e. orange line in the above panels, but it is able to purify the input signal also for other pump durations.  

\begin{figure}[h!]
    \includegraphics[width=\linewidth]{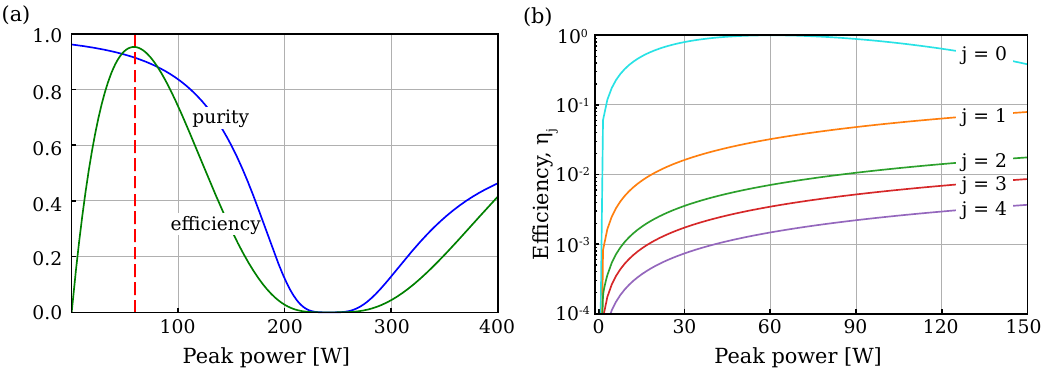}
    \caption{Applying our scheme with a noisy QD photon. 
    (a) Conversion efficiency of the zeroth mode normalised with respect to all the Schmidt modes of the state and corresponding output purity.
    The red dashed line highlights the maximum conversion efficiency point $\sim$96\% at a peak power of $60$~W, closely approaching ideal single-mode conversion.
    (b) Unnormalized conversion efficiency of the first five Schmidt modes.
    As the peak power increases beyond unit conversion efficiency in the zeroth mode, higher-order contributions increase.
    }
    \label{Fig: efficiency_and_purity}
\end{figure}

To exemplify our scheme, as shown in \cref{fig:purity_with_freq_time_noise}, we will analysise its performance for a QD with spectral purity $\sim$~0.76 as reported in ~\cite{Somachi_2016}.
This corresponds to inhomogenous broadening of $\sim$~1~GHz for a QD emitting photons with transform-limited linewidth of 1~GHz, assuming the noise is only in the spectral domain---a similar treatment could be realised for temporal noise.
In \cref{Fig: efficiency_and_purity}(a) we plot output state purity as a function of peak power, with a waveguide length of $L=15$~mm and pump pulse duration of 13~ps.
We also plot the normalized conversion efficiency of the zeroth mode, $\eta_0$, as a function of peak power.
The normalization is performed over all Schmidt modes of the output photon, conditional on successful frequency conversion.
While the purity of the output photon is highest as the peak power tends to zero, in this regime the normalized efficiency of the zeroth mode also tends to zero and is not practical.
Instead, we observe the maximum normalized conversion efficiency of the zeroth mode approaches 96\% with 60~W peak power, although the overall purity falls to 91.8\%.
Prima facie, the maximum normalized zeroth mode conversion efficiency represents a good configuration as this leads to the highest likelihood of frequency conversion into the desired spectral mode.
However, if the purity of the overall state has an outsized importance then the pump power can be lowered to increase the purity at the cost of efficiency.

By plotting the unnormalized conversion efficiencies of the first five Schmidt modes as a function of peak power \cref{Fig: efficiency_and_purity}(b) we explicitly show how the final purity trade-off with efficiency of conversion into the zeroth mode behaves.
We observe the efficiency of the zeroth mode reaches its maximum, i.e., unit conversion efficiency when un-normalized, at lower peak powers than higher order modes.
Operating beyond this peak power will decrease the purity as the higher-order modes are also converted, and the zeroth mode begins to decrease.

An important point to consider is whether our scheme outperforms passive filtering, such as a Fabry-Perot, which is typicall used to ensure a narrow spectrum of the photon emitted by a QD~\cite{Thomas2024sciadv,Kambs_16}.
We benchmark our purification scheme against using intensity filters by comparing the transmission efficiency for both of the two techniques.
Following~\cite{gao2019physrevlett}, our simulated transmission efficiency reaches $\sim$~96\%, in principle exceeding passive filtering, with similar bandwidths, which reached $\sim$~70\%~\cite{Somachi_2016}.
Furthermore, the passive filtering technique cannot address timing jitter~\cite{Raymer_2020}, given that by definition it is time-stationary.
This comparison underscores the advantages of our technique in both transmission efficiency and noise reduction.

So far, we have only considered spectral purification using FC in the counter-propagating configuration of the input and output fields, however spectral purification can also be achieved in the co-propagating regime.
To establish a comparison with the counter-propagating configuration, we adopt the same waveguide length in the previous example, e.g., $L$=15~mm, and we optimize the pulse duration to achieve the same level of purity.
The immediate difference is that the output photon bandwidth is 1.61~THz, which represents three orders of magnitude increase. 
This highlights the key advantage of considering the counter-propagating phase matching condition for preserving narrow spectral profile of the single photon.
However this method requires short poling features which is currently technologically demanding.
Sub-micron poling periods have been demonstrated in the context of parametric down-conversion sources using potassium titanyl phosphate (KTP) bulk crystals~\cite{Kuo2023opticaquantum} and LN thin-film waveguides~\cite{Nagy2020OME,SLAUTIN2021CerInt}.
Nonetheless, achieving this level of precision in a scalable fashion remains a significant technological challenge.
One potential remedy to this issue is to increase the quasi-phase matching order $m$, albeit at the cost of reduced conversion efficiency, which scales down as $1/m^2$.
For example, setting $m=3$ for our PPLN example would require a poling period of 1 $\mu \text{m}$ and yield a peak efficiency of approximately 10\% while maintaining the same purity level at the same peak power.
Although this approach significantly reduces maximum efficiency, it makes the technique technologically more feasible.

Note that in this work we have neglected the time-ordering effects which have been investigated in sum frequency generation configuration where it has been shown to limit the peak conversion efficiency to $\sim$~90\%\cite{Christ_2013}.
This effect is typically only observed in the high-gain nonlinearity regime, where the pump field intensity is sufficiently large, or equivalently when considering frequency conversion processes near their maximum efficiencies.
As such a similar treatment for our scheme would be necessary to establish the absolute upper limits of the conversion efficiency, we leave this for future work.

\section{Conclusions}

We introduced a coherent quantum interface which spectrally purifies an input signal during conversion to telecom wavelength without broadening the typical GHz-range input bandwidth of QD single-photon sources.
Quantum FC is commonly used to convert QD photons from $\sim$900~nm to the telecom band~\cite{pelc2012,Kambs_16,Weber_2018,Weber_2019, morrison2021apl,da2022,morrison2023single}, often using a CW pump which can reach unit internal conversion efficiency but retains spectral noise~\cite{Brecht_2011}.
By using a pulsed pump, we can optimize the conversion efficiency for desired modes~\cite{brecht2015physrevx}, resulting in near-perfect single-mode conversion that leads to high spectral purity and indistinguishability, which is essential for advanced quantum communication protocols.
Furthermore, we exploit the counter-propagating configuration to ensure narrow spectral bandwidth of the output photon after conversion.
A counter-propagating FC scheme has been demonstrated by \emph{Guo et al.} which primarily sought to achieve spectral compression for conversion from 550~nm to 1545~nm~\cite{Guo2021FrontPhys}.
Our work is thus related in that we also exploit this configuration, however we further expand on this by introducing the spectral purification for InGaAs QD and its conversion to telecom wavelength.
Additionally, by optimizing parameters such as the pump pulse duration and waveguide length, our scheme can produce narrower spectral bandwidths of the converted photon compared to~\cite{Guo2021FrontPhys}, bringing us closer to match the wavelengths and bandwidths of different quantum platforms including broadband quantum memories~\cite{sprague2014broadband,Hepp2014PRL,Saunders2016PRL}.

A crucial aspect for integrating single quantum emitters into quantum networks is their operational feasibility outside laboratory environments.
For QDs, this is challenging owing to the large and power-demanding cryogenic cooling systems that are typically required to ensure low-noise operation~\cite{yu2023naturenano}.
By operating at higher temperatures, the size, weight and power requirements can be decreased, however it comes at the cost of spectral purity and indistinguishability~\cite{Thoma_2016,Gerhardt_2018,Brash_2023}.
As such, our technique which operates at room temperature complement this effort to designing compact and high-performing single photon sources at telecom wavelengths.

Here we have presented the specific case where aGVM was shown for FC from 942~nm to 1550~nm, in addition we have confirmed that the aGVM condition holds over the broader range of conversion from 920-980~nm, typical for InGaAs semiconductor QDs, to the 1500-1600~nm.
As such, our scheme can be adapted by selecting the appropriate pump parameters to match the specific waveguide length to implement the spectral purification technique for QDs emitting within this range.
Additionally, the scheme for LN also works in reverse as such allowing purification through FC from telecom C-band wavelengths to shorter wavelengths to enable interfacing with quantum memories from relatively noisy quantum dots emitting at telecom wavelengths~\cite{Thomas2024sciadv}

A natural extension of our work is to consider purifying entangled states directly generated by a single QD through biexciton-exciton cascade radiative processes~\cite{yu2023naturenano,Liu_2019} to enable entanglement generation for advanced quantum networking applications.
Going beyond QDs, there are a range of other non-telecom single photon emitters for which FC to telecom wavelengths has been demonstrated~\cite{Matthias_2018,Anais_2018,mann2023}.
As such, our technique can be applied to other single quantum emitters to purify the converted photons by identifying the respective aGVM configurations.

\ack

The authors want to thank N. Quesada and F.B. Barnes for helpful discussion and feedback on the manuscript.

\section*{References}

\bibliographystyle{unsrt}
\bibliography{Main}

\end{document}